\documentstyle[twoside,fleqn,espcrc2,psfig]{article}
\newcommand{\be}{\begin{eqnarray}}
\newcommand{\ee}{\end{eqnarray}}
\newcommand{\la}{\langle}
\newcommand{\ra}{\rangle}
\newcommand{\no}{\nonumber}
\newcommand{\Op}{{\cal O}}
\title{A lattice NRQCD computation of the bag parameters for
       $\Delta B = 2$ operators\thanks{Presented by N.Yamada.}
      }
\author{
  N. Yamada\address{
    Department of Physics, Hiroshima University,
    Higashi-Hiroshima 739-8526, Japan
    },
  S. Hashimoto\address{
    High Energy Accelerator Research Organization(KEK),
    Tsukuba 305-0801, Japan
    },
  K-I. Ishikawa$^{\rm b}$,
  H. Matsufuru$^{\rm a}$ and
  T. Onogi$^{\rm a}$
  }
\begin{document}
\begin{abstract}
  We present an update of our NRQCD calculation of $B_B$ at
  $\beta$=5.9 with increased statistics. 
  We also discuss a calculation of $B_S$, which is relevant 
  to the width difference in the $B_s-\bar{B}_s$ mixing.
\vspace{-80mm}
\begin{flushright}
\large HUPD-9917
\end{flushright}
\vspace{70mm}
\end{abstract}
\maketitle
\section{Introduction}

The NRQCD calculation is essential to obtain a prediction
for $B_B$ with precision better than $O(20\%)$, as the size
of $1/M$ correction, which is not included in the static
calculations, is expected to be $\Lambda_{QCD}/m_b$ = 
0.1$\sim$0.2. 
We update our study of $B_B$ using the NRQCD, which was
previously presented at the last lattice conference
\cite{yamada_lat98}. 
A paper version is also available \cite{yamada_paper}.

Based on the same calculation method, we have also calculated 
$B_S$, which is relevant to the width difference of $B_s$
meson system \cite{beneke}.

\section{Method}

The bag parameter $B_{X_q}(m_b)$ is defined using a vacuum
saturation approximation (VSA) as
\be
    B_{X_q}(m_b)
    &=& \frac{\la\bar{B}_q^0|\Op_{X_q}(m_b)|B_q^0\ra}
              {\la\bar{B}_q^0|\Op_{X_q}(m_b)|B_q^0\ra_{\rm VSA}}
\label{eq1},
\ee
where the $\Delta B$=2 operators $\Op_{X_q}$ are
$\Op_{L_q}=\bar{b}\gamma_\mu P_L q\ \bar{b}\gamma_\mu P_L q$ 
or $\Op_{S_q}=\bar{b} P_L q\ \bar{b} P_L q$.
($P_L$ is a projection operator $P_L=1-\gamma_5$.)
Subscript $q$ denotes the valence light quark $d$ or
$s$, which we omit in the following if there is no risk of
confusion. 
We use a notation $B_L$ instead of $B_B$ to remind that it
is a matrix element of $\Op_L$ and to distinguish it from
$\Op_S$. 

Using the operators constructed on the lattice 
with static heavy and clover light quark $O_X^{\rm lat}$, 
the continuum operators defined with the $\overline{MS}$
scheme $\Op_X$ are written as
\be
    \Op_L(m_b)
&=& \hspace{-5mm}\sum_{X=\{L,S,R,N\}} Z_{L,X} O_X^{\rm lat}(a^{-1}),\\
    \Op_S(m_b)
&=& \hspace{-5mm}\sum_{X=\{S,L,R,P\}} Z_{S,X} O_X^{\rm lat}(a^{-1}),\\
    {\cal A}_0
&=& Z_A A_0^{\rm lat},
\ee
where new operators $\Op_R$, $\Op_N$ and $\Op_P$ are involved:
\be
        \Op_R
&=&     \bar{b}\gamma_\mu P_R q\ \bar{b}\gamma_\mu P_{R} q,\no\\
        \Op_N
&=&  2\ \bar{b}\gamma_\mu P_L q\ \bar{b}\gamma_\mu P_R q
   + 4\ \bar{b}P_Lq\ \bar{b} P_R q,        \label{dif_op}\no\\
        \Op_P
&=&   2\ \bar{b}\gamma_\mu P_Lq\ \bar{b}\gamma_\mu P_R q
   + 12\ \bar{b}P_Lq\ \bar{b} P_R q.           \no
\ee
$Z_{L,X}$ and $Z_{S,X}$ are perturbative matching
factors obtained at one-loop level \cite{matching}.
We also write the matching of the heavy-light axial current 
${\cal A}$ with the renormalization constant $Z_A$.

The bag parameters are, then, written in terms of 
the corresponding quantities measured on the lattice
$B_{X}^{\rm lat}$ as 
\be
    B_L(m_b)
&=& \hspace{-6mm}\sum_{X=\{L,S,R,N\}}\hspace{-3mm}
    Z_{L,X/A^2} B_X^{\rm lat}(a^{-1}),\\
\label{eq4}
    B_S(m_b)/0.734
&=& \hspace{-6mm}\sum_{X=\{S,L,R,P\}}\hspace{-3mm}
    Z_{S,X/A^2} B_X^{\rm lat}(a^{-1}).
\ee
Here $Z_{L,X/A^2}$ denotes a ratio of matching constants
$Z_{L,X}/Z_A^2$, and $B_X^{\rm lat}$ is defined by
\be
    B_X^{\rm lat}(a^{-1})
&=& \frac{\la\bar{B}^0|\Op^{\rm lat}_X(a^{-1})|B^0\ra}
    { c
      \la\bar{B}^0| A_\mu^{\rm lat} |0\ra
      \la 0| A_\mu^{\rm lat} |B^0\ra}.
\ee
A numerical constant $c$ is 8/3 or $-$5/3 in $B_L$ or in $B_S$ 
respectively. 

The vacuum saturation of the operator $\Op_S$ introduces a
matrix element of the pseudoscalar density
$P=\bar{b}\gamma_5 q$, which is often rewritten in terms of
$A_{\mu}$ using the equation of motion.
In doing so, a factor 
$(\overline{m}_b(m_b)+\overline{m}_s(m_b))^2/M_{B_s}^2$
appears, for which we use 
\be
m_b=4.8\ {\rm GeV},&&\ \bar{m_b}(m_b)=4.4\ {\rm GeV},\no\\
\bar{m_s}(m_b)=0.2\ {\rm GeV},&&\ M_{B_s}=5.37\ {\rm GeV}\no
\ee
as in Ref.\cite{beneke}, and obtain 0.734 given in
Eq.(\ref{eq4}). 

Unfortunately the one-loop coefficients for the perturbative 
matching are not yet available for the NRQCD action.
We use, therefore, the one-loop coefficients calculated in
the static limit as an approximation.
It introduces a systematic error of $O(\alpha_s/(am_Q))$,
but no logarithmic divergence appears.
The numerical values of $Z_{L,X/A^2}$ and $Z_{S,X/A^2}$ at
$\beta$=5.9 are given in Table \ref{matchfac}, in which we
linearize the perturbative expansion of $Z_{L,X}/Z_A^2$ and
neglect all the $O(\alpha_s^2)$ terms. 
For the coupling constant, $\alpha_V(q^*)$ with $q^*=1/a$
and $\pi/a$ is used throughout this paper.
\begin{table}[t]
\vspace{-0.5cm}
\begin{center}
\begin{tabular}{|c||c|c|c|c|}
\hline
$q^*$ & $Z_{L,L/A^2}$ & $Z_{L,S/A^2}$ & $Z_{L,R/A^2}$ & $Z_{L,N/A^2}$\\
\hline\hline
$\pi/a$ & 0.973 & -0.104 & -0.007 & -0.080 \\
$1/a$   & 0.956 & -0.172 & -0.011 & -0.132 \\
\hline\hline
$q^*$ & $Z_{S,S/A^2}$ & $Z_{S,L/A^2}$ & $Z_{S,R/A^2}$ & $Z_{S,P/A^2}$\\
\hline\hline
$\pi/a$ & 1.307 & 0.032 & 0.002 & 0.010 \\
$1/a$   & 1.505 & 0.053 & 0.003 & 0.017 \\
\hline
\end{tabular}
\label{matchfac}
\caption{Perturbative matching factors at $\beta$=5.9.}
\vspace{-9mm}
\end{center}
\end{table}

Our simulation was carried out on a quenched 
$16^3 \times 48$ lattice at $\beta$=5.9.
We have increased the statistics to 250 from 100 at the time 
of Lattice 98 \cite{yamada_lat98}.
We performed two sets of simulations with the NRQCD actions
and currents improved through $O(1/m_Q)$ and $O(1/m_Q^2)$,
which enables us to study the higher order effects in the
$1/m_Q$ expansion explicitly. 
The light quark is described by the clover action with the
tadpole improved clover coefficient $c_{sw}=1/u_0^3$.
The inverse lattice spacing is determined from the rho meson
mass as $a^{-1}$ = 1.62 GeV.

\section{$B_L$}

\begin{figure}[t]
\leavevmode\psfig{file=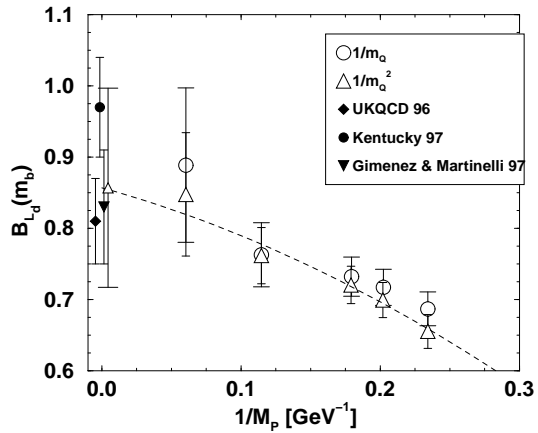,width=7cm}
\vspace{-15mm}\\
\caption{The heavy quark mass dependence of $B_{L_d}(m_b)$.}
\label{fig1}
\vspace{-5mm}
\end{figure}

Figure \ref{fig1} shows $1/M_P$ dependence of
$B_{L_d}(m_b)$ ($q^*$=$1/a$) with $M_P$ the pseudoscalar
heavy-light meson mass. 
Open circles denote the results with the $O(1/m_Q)$ NRQCD
action and open triangles denote those with the $O(1/m_Q^2)$
action.

We fit the $O(1/m_Q^2)$ results to a quadratic function of
$1/M_P$ (dashed line) and obtain the value in the static
limit (small open triangle).
We also plot the previous results in the static limit by 
UKQCD \cite{ukqcd96} (filled diamond), Kentucky group
\cite{cdm} (filled circle) and Gim\'enez and
Martinelli~\cite{GM} (filled triangle).
In order to make a consistent comparison we reanalyzed their 
data using the same matching procedure described in the last
section. 
Our data extrapolated to the static limit nicely agrees with
these direct simulation results, as it should be.

From Figure~\ref{fig1} we observe that $B_L$ has a small
negative slope in $1/M_P$, which is well described by the
vacuum saturation approximation
\cite{yamada_lat98,yamada_paper} and also observed in the
lattice calculations with relativistic actions
\cite{draper,shoji}. 
We also find that the $O(1/m_Q^2)$ corrections to the action 
and current gives only a few per cent contribution to $B_L$.

The dominant uncertainty in our result comes from the
unknown one-loop coefficients for the NRQCD action.
A crude estimate with order counting suggests that the
corresponding systematic error is $O(\alpha_s/(am_b))$
$\sim$ 10\%. 
Other possible systematic errors are the discretization
error of $O(a^2\Lambda_{QCD}^2)$ and of 
$O(\alpha_s a\Lambda_{QCD})$, the relativistic correction of 
$O(\Lambda_{QCD}^2/m_b^2)$, and a small uncertainty in the
chiral extrapolation.

Taking them into account, we obtain the following values as 
our final results from the quenched lattice,
\be
  B_{B_d}(m_b) = 0.75(3)(12),\ \
  \frac{B_{B_s}}{B_{B_d}} = 1.01(1)(3),
\ee
where the first error is statistical and the second a sum of
the systematic errors in quadrature. 
In estimating the error in the ratio $B_{B_s}/B_{B_d}$ we
consider the error from chiral extrapolation only, assuming
that other uncertainties cancel in the ratio.

\section{$B_S$}

\begin{figure}[t]
\leavevmode\psfig{file=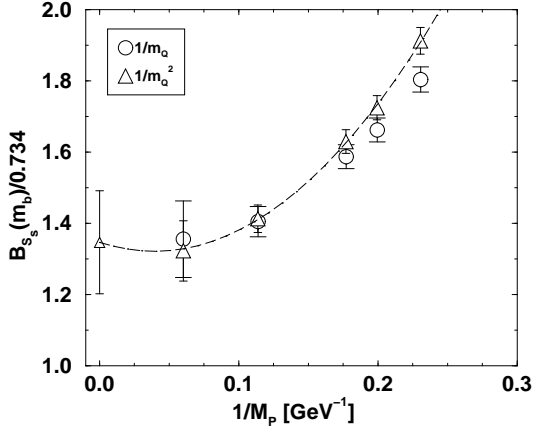,width=7cm}
\caption{The heavy quark mass dependence of $B_{S_s}(m_b)$.}
\label{fig2}
\end{figure}

Figure \ref{fig2} shows the $1/M_P$ dependence of
$B_{S_s}(m_b)$ with $q^*=1/a$.
We see a significant increase of $B_S$ with the $1/M$
correction, which is 20$\sim$30\%.
Our preliminary result with a similar error analysis as in
$B_L$ is
\be
  B_{S_s}(m_b) = 1.19(2)(20).
\ee

The width difference in the $B_s-\bar{B}_s$ mixing
$\Delta\Gamma_s$ is theoretically calculated using the $1/M$
expansion as \cite{beneke}
\be
\lefteqn{
  \left(\frac{\Delta\Gamma}{\Gamma}\right)_s
  =
 \left(\frac{f_{B_s}}{210 {\rm MeV}} \right)^2 }\no\\
&\times& \left[  0.006\ B_{L_s}(m_b)
             + 0.150\ B_{S_s}(m_b)
             - 0.063 \right].\no
\ee
Using our result for $B_L$ and $B_S$, and a recent dynamical 
lattice result $f_{B_s}$ = 245(30)~MeV \cite{shoji}, we
obtain $(\Delta\Gamma/\Gamma)_s$ = 0.16(3)(4), where errors
are from $f_{B_s}$ and from $B_S$ respectively.

\vspace{7mm}
Numerical calculations have been done on Paragon XP/S at
Institute for Numerical Simulations and Applied Mathematics
in Hiroshima University.
We are grateful to S. Hioki for allowing us to use his program
to generate gauge configurations.
S.H. and T.O. are supported in part by the Grants-in-Aid of
the Ministry of Education (Nos. 11740162,10740125).
K-I.I. would like to thank the JSPS for Young Scientists
for a research fellowship.


\end{document}